\documentclass[aps, pre, twocolumn, groupedaddress, showpacs, 10pt]{revtex4-1}
\usepackage{graphicx}
\usepackage[toc,page]{appendix}
\usepackage[latin1]{inputenc}
\usepackage{amsmath, amsfonts, amssymb, makeidx, graphicx, float, color, url, setspace, subfigure, rotating}
\usepackage[pdftex, bookmarks, colorlinks, breaklinks]{hyperref}\hypersetup{linkcolor = red, citecolor = blue, filecolor = black, urlcolor = blue}
\hypersetup{pdfauthor = {Chiranjit Mitra},
pdftitle = {Report}, pdfsubject = {Multi-scale dynamics of glow discharge plasma through wavelets: Self-similar behavior to neutral turbulence and dissipation}}
\usepackage[section]{placeins}

\begin{document}

\title{Multi-scale dynamics of glow discharge plasma through wavelets: Self-similar behavior to neutral turbulence and dissipation}

\author{Bapun K. Giri}
\email[]{bapunk@iiserkol.ac.in}
\affiliation{Department of Physical Sciences, Indian Institute of Science Education and Research (IISER) Kolkata, Mohanpur, 741252, India.}

\author{Chiranjit Mitra}
\email[]{chiranjitmitra4u@iiserkol.ac.in}
\affiliation{Department of Physical Sciences, Indian Institute of Science Education and Research (IISER) Kolkata, Mohanpur, 741252, India.}

\author{Prasanta K. Panigrahi}
\email[]{pprasanta@iiserkol.ac.in}
\affiliation{Department of Physical Sciences, Indian Institute of Science Education and Research (IISER) Kolkata, Mohanpur, 741252, India.}

\author{A. N. Sekar Iyengar}
\email[]{ansekar.iyengar@saha.ac.in}
\affiliation{Plasma Physics Division, Saha Institute of Nuclear Physics (SINP) Kolkata,
 1/AF, Bidhannagar, Kolkata 700064, India.}



\begin{abstract}
The multiscale dynamics of glow discharge plasma is analysed through wavelet transform, whose scale dependent variable window size aptly captures both transients and  non-stationary periodic behavior. The optimal time-frequency localization ability of the continuous Morlet wavelet is found to identify the scale dependent periodic modulations efficiently, as also the emergence of neutral turbulence and dissipation, whereas the discrete Daubechies basis set has been used for detrending the temporal behavior to reveal the multi-fractality of the underlying dynamics. The scaling exponents and the Hurst exponent have been estimated through wavelet based detrended fluctuation  analysis,and also Fourier methods and rescale range analysis.
\end{abstract}

\maketitle


\section{Introduction}

Glow discharge plasma \cite{Shukla} is a prototypical nonlinear dynamical system and an ideal test bed for observing a rich variety of nonlinear phenomena e.g., chaos, frequency entrainment \cite{Frequency_entrainment}, intermittency \cite{Intermittency} and turbulence \cite{Plasma_Review,Jha}. These have been investigated by nonlinear time series analysis, spectral analysis and other methods. Turbulence and fluctuations in plasma play crucial role in the transport of charge and energy, which makes their investigation an area of active study \cite{Turbulence}. However, established techniques of nonlinear time series analysis, such as correlation dimension, Lyapunov exponent, fractal dimension \cite{Mandelbrot,Feder} only probe the global and asymptotic properties of the system and are ineffective for unravelling the local structures. Some of these techniques limit themselves to the spectral amplitude of the signal, neglecting the phase information \cite{Nurrujaman}. Furthermore, in many of these approaches the short time scale characteristics are often overlooked, making it difficult to obtain the time history and local properties of the system.

We report here, results of a wavelet based analysis of floating potential fluctuations in an argon DC glow discharge plasma in a magnetic field. Being a local approach, it brings out the non-stationary nature of this dynamical system quite effectively \cite{WA_Book}. We make use of both discrete and continuous wavelets to probe different aspects of the dynamics. Using discrete wavelets belonging to the Daubechies family \cite{Daubechies}, we have isolated the local polynomial trends and have quantitatively estimated the multi-fractal characteristics \cite{WA_Stanley,WA_Panigrahi_1,WA_Panigrahi_2,Morlet,Manimaran}. The continuous Morlet wavelet \cite{Morlet,Morlet_2} economically identifies the periodic and structured variations in the time-frequency domain, whose nature is probed as a function of magnetic field on the floating potential.

The paper is organized as follows: Section \ref{sec:Setup} provides a brief outline of the experimental setup, after which section \ref{sec:FA} details the results from Fourier analysis, which reveals the periodic components and also the self-similar behavior of the dynamical system. However, it is clearly inadequate to throw light on the time-varying structures and also the multi-fractality of the data. Section \ref{sec:WA} uses wavelet transform to precisely capture the above dynamical features of the system. Apart from time-frequency localization of the periodic modulations, wavelet transform helps unravel phase synchronization among different modes. The continuous Morlet wavelet clearly identifies the parameter domain where the dynamical system displays neutral turbulence and dissipation, obeying Heisenberg model. Finally, we conclude in section \ref{sec:Conclusion}, after outlining the complementary nature of different methods employed and highlighting the usefulness of local approaches. The directions of further studies are also laid.


\section{Experimental Setup} \label{sec:Setup}

\begin{figure}[!]
\centering
\includegraphics[trim = 120mm 0mm 0mm 0mm, clip,scale=0.35]{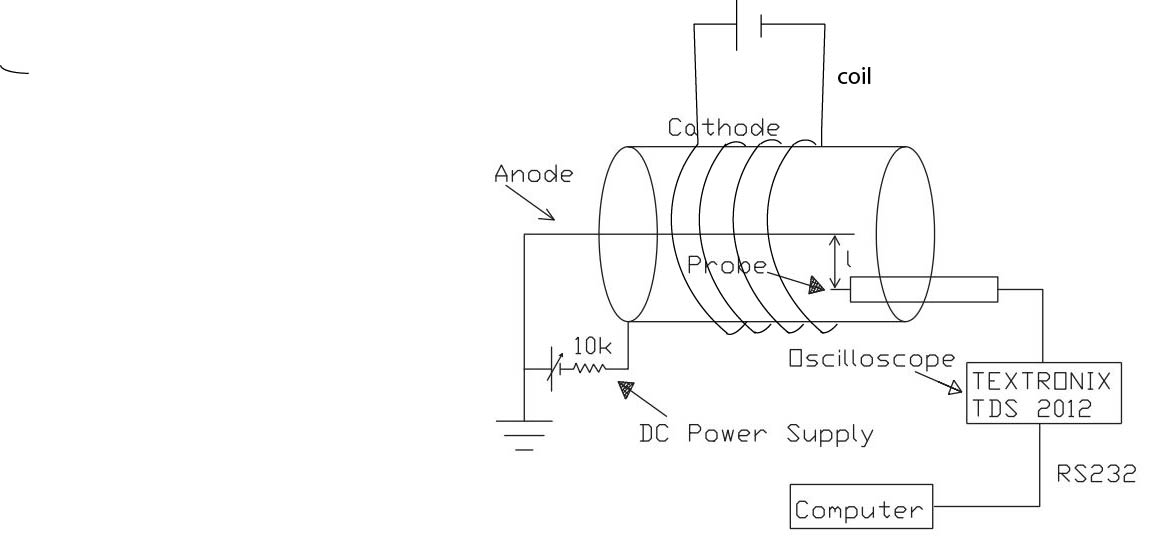}
\caption{\label{fig:Setup}Schematic diagram of the experimental setup.}
\end{figure}

The experiments were performed in a hollow cathode DC glow discharge plasma device, which is schematically shown in figure \ref{fig:Setup}. It comprises of a hollow cathode tube with a central wire and a Langmuir probe for monitoring the complex temporal dynamics in the plasma floating potential time series. This assembly was mounted inside a vacuum chamber and pumped down to a pressure of about $\sim$0.001 mbar and subsequently filled with argon gas up to a pressure $\sim$0.077 mbar. Plasma was formed at $\sim$330 V by a DC discharge voltage power supply (which could be varied from 0 to 1000 V). A copper wire was wrapped around the cathode externally. By passing current through the copper wire, magnetic field was generated inside the cathode tube. Hence, there are three control parameters, namely pressure, discharge voltage and applied magnetic field. In the present case, pressure and electrode configuration were kept constant throughout the experiment.

The floating potential time series obtained through the Langmuir probe is the object of study here. For this purpose, the data at two different DC discharge voltages, 330 V and 430 V have been obtained. At a particular DC discharge voltage, the applied magnetic field was varied by changing the current in the copper wire (outside coil). The fluctuations of the plasma floating potential were recorded through a digital oscilloscope. Table \ref{Mag} shows the DC discharge voltages and the corresponding magnetic field parameter used for the present experiment.

\begin{table}[h]
\caption {Applied Discharge Potential and corresponding magnetic fields } \label{Mag} 

\begin{tabular}{|c|c|}
\hline 
Potential(V) & Magnetic Field (gauss) \\ \hline \hline
 & 0 \\
 &  19 \\ 
 & 28 \\
330 & 37 \\
 & 46 \\
 & 87 \\
 & 96 \\ \hline
 & 0 \\	
    & 9 \\
   	& 19 \\             
    & 28 \\
430 & 37 \\
    & 46 \\
    & 57 \\
    & 67 \\ \hline

\end{tabular}
\end{table}

\FloatBarrier

\begin{figure}[h]
    \centering
    \subfigure[]
    {
        \includegraphics[scale=0.3]{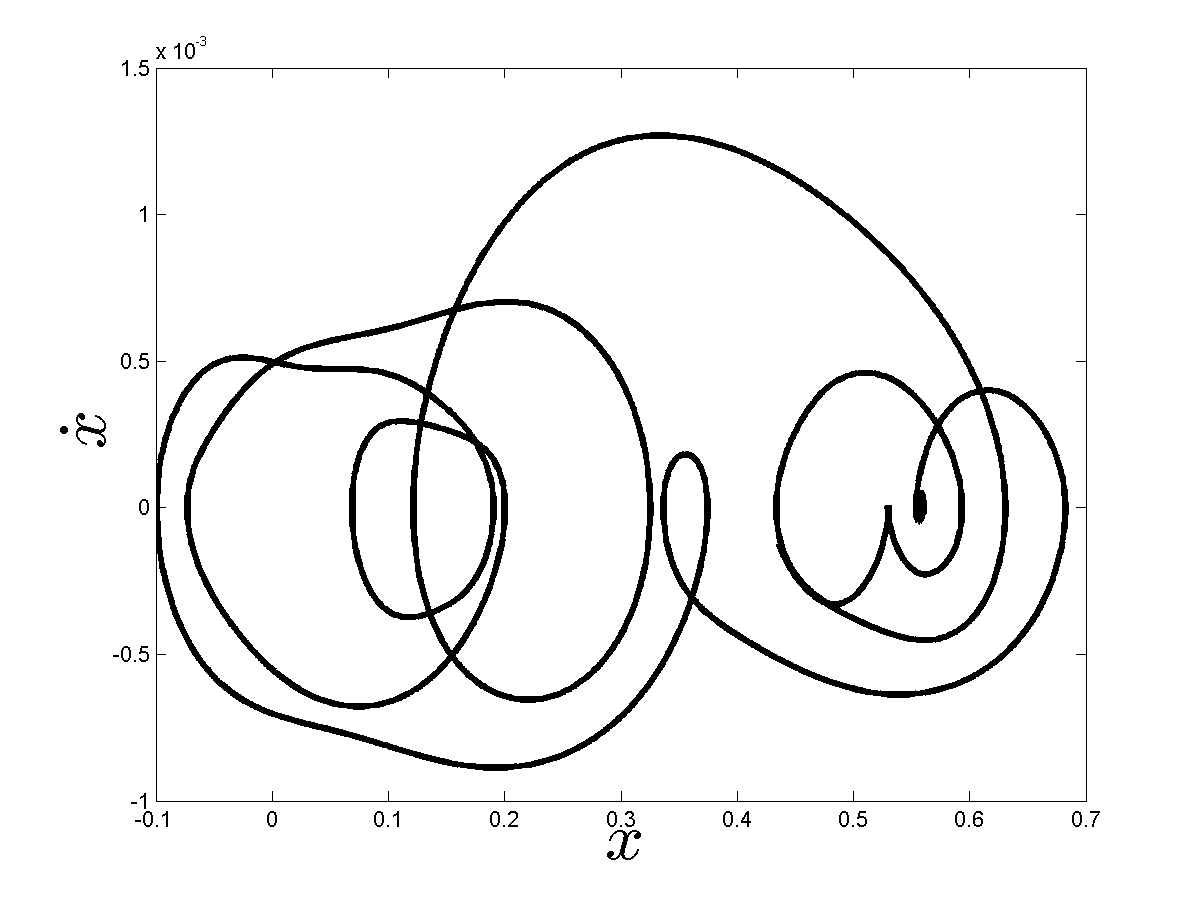}
        \label{fig:Phase_1}
    }
    \\
    \subfigure[]
    {
        \includegraphics[scale=0.3]{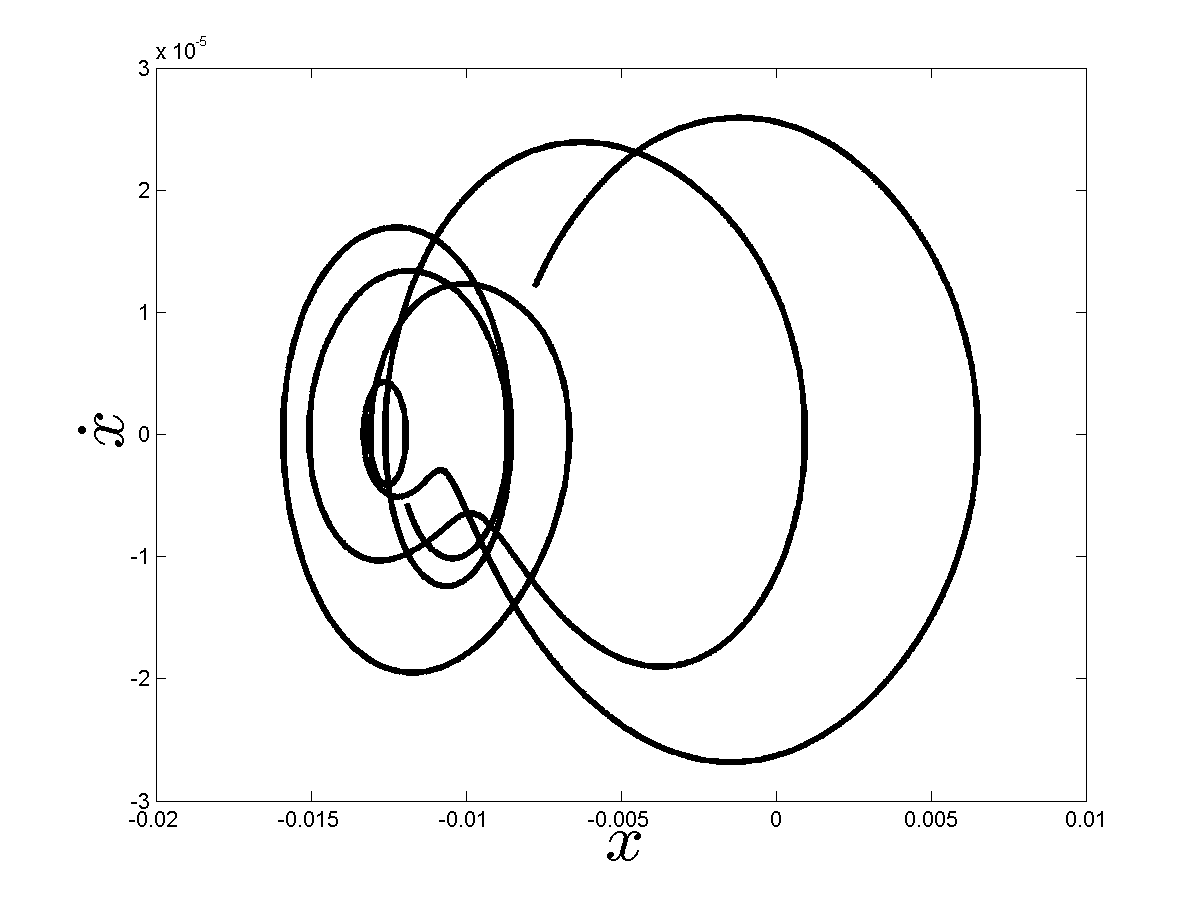}
        \label{fig:Phase_2}
    }
    \caption{Phase space plots of two signals ($x(t)$) at (a) 330 V, 96 gauss and (b) 430 V, 19 gauss respectively, illustrating the periodic dynamics.}
\end{figure}

Figures \ref{fig:Phase_1} and \ref{fig:Phase_2} depict the phase space plot of two signals at 330 V, 96 gauss and 430 V, 19 gauss respectively, obtained by filtering out the high frequency components. These plots illustrate the periodic behavior of the signals.

\FloatBarrier

We now proceed to study the Fourier domain behavior of the floating potential fluctuations in the time series to show the presence of both periodic, as well as self-similar behavior.


\section{Fourier Analysis} \label{sec:FA}

\begin{figure}[!]
\begin{center}$
\begin{array}{cc}
\includegraphics[width=3in]{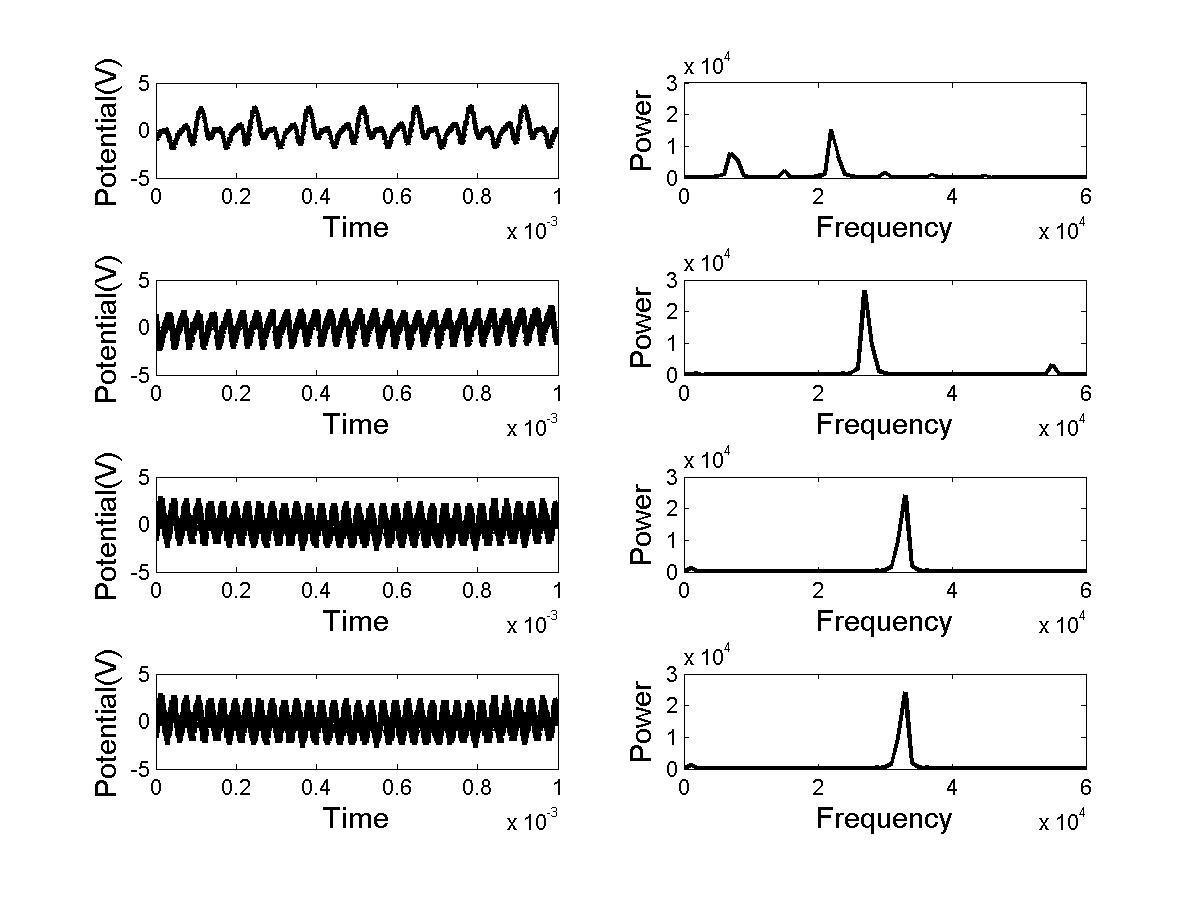} \\
\includegraphics[width=3in]{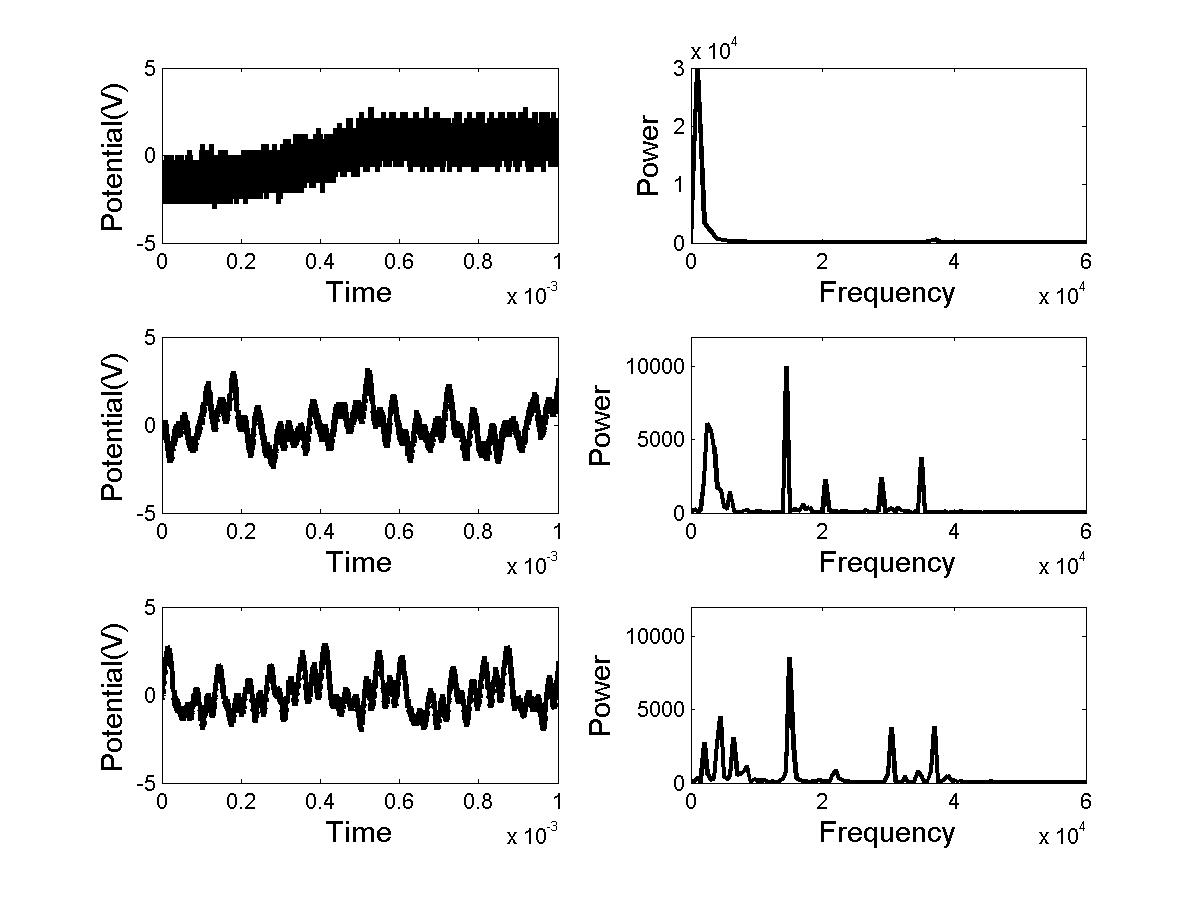}
\end{array}$
\end{center}
\caption{\label{fig:FFT_1}Signals at 330 V and their respective power spectra, as the magnetic field is increased.}
\end{figure}

\begin{figure}[!]
\begin{center}$
\begin{array}{cc}
 \includegraphics[width=3in]{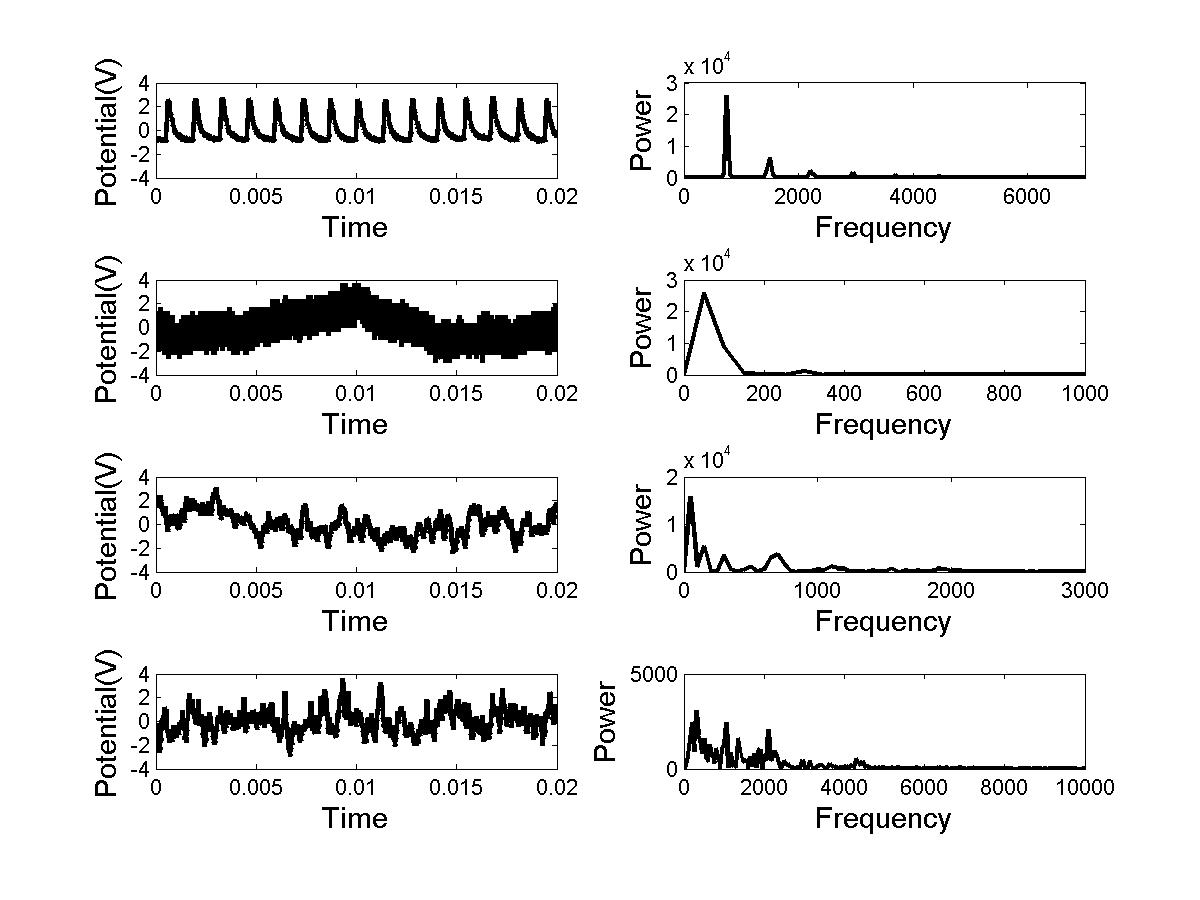} \\
\includegraphics[width=3in]{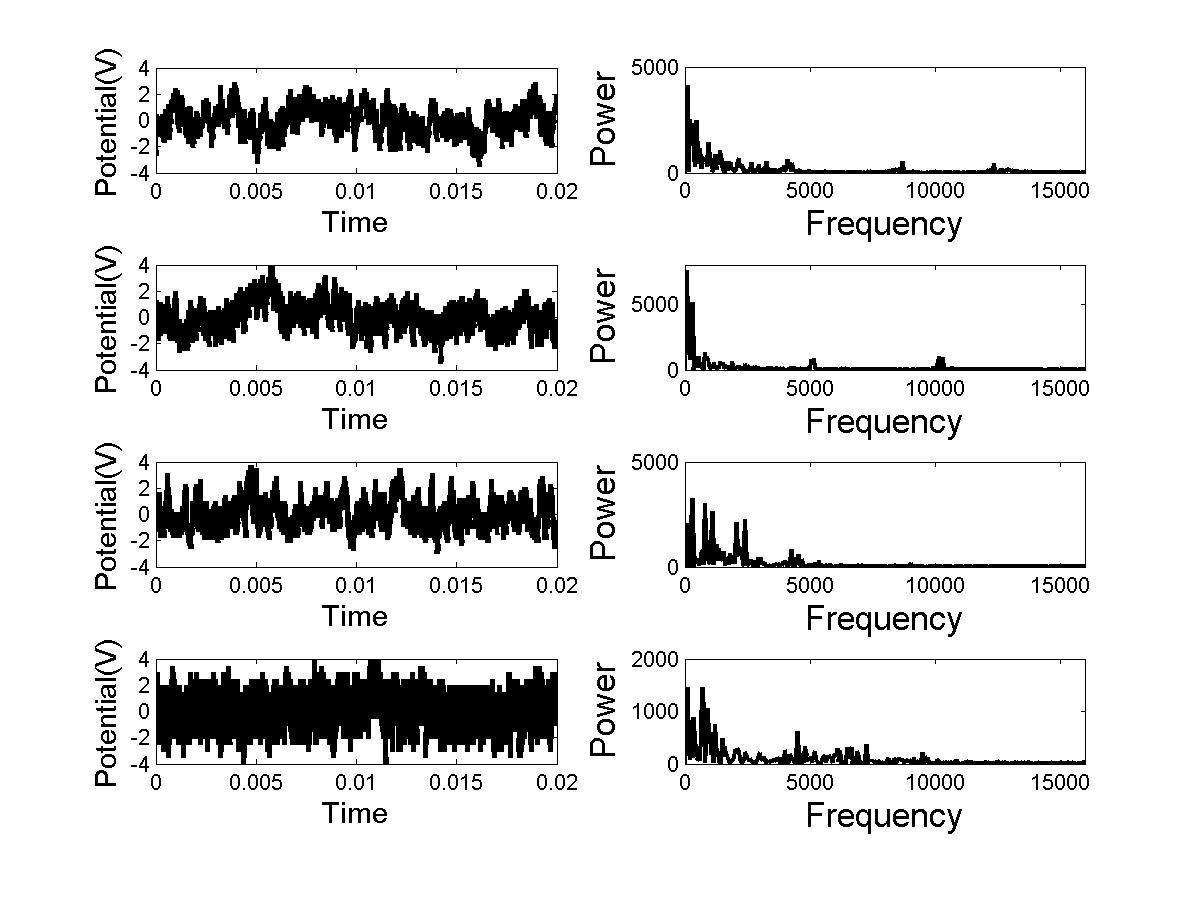}
\end{array}$
\end{center}
\caption{\label{fig:FFT_2}Signals at 440 V and their respective power spectra, as the magnetic field is increased.}
\end{figure}

\begin{figure}[!]
    \centering
    \subfigure[]
    {
        \includegraphics[height=1.8in,width=3.5in]{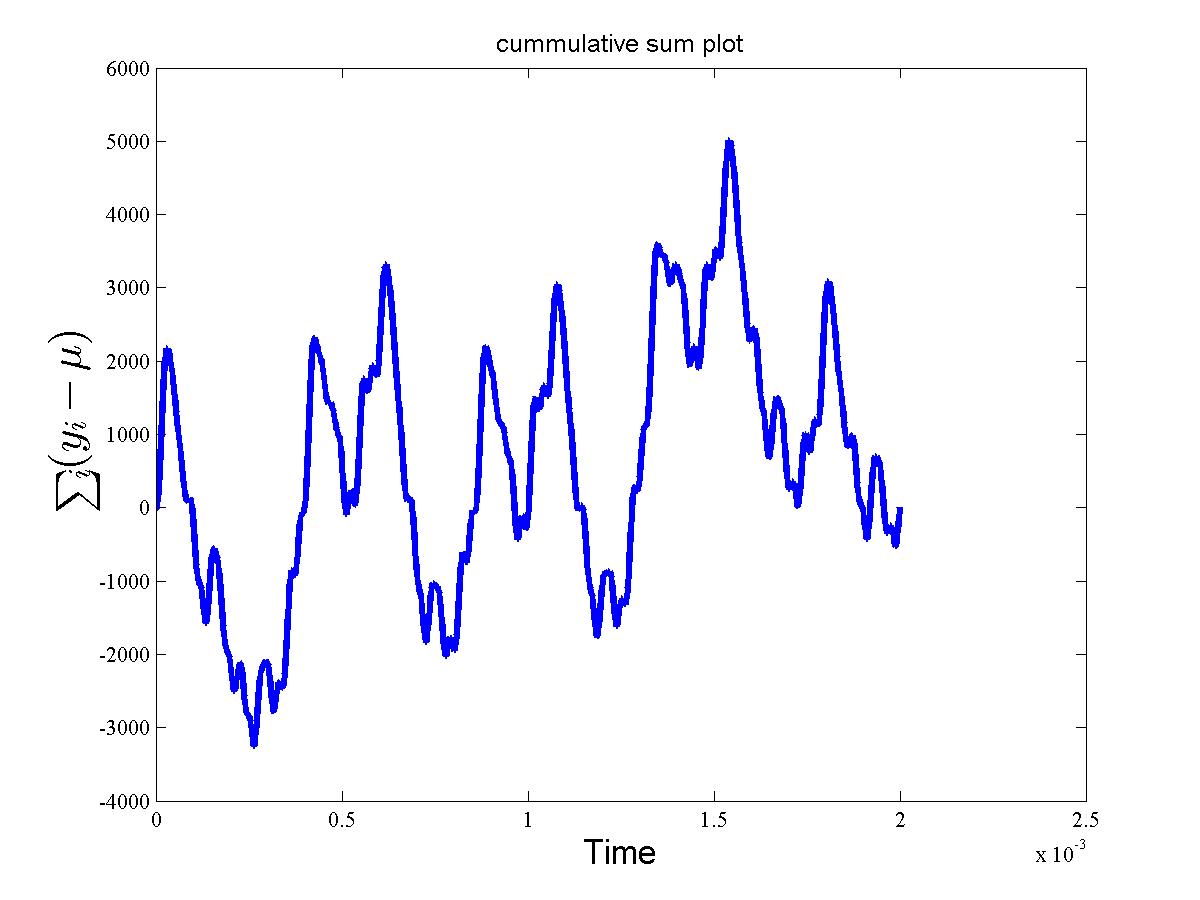}        
    }
    \\
    \subfigure[]
    {
        \includegraphics[height=1.8in,width=3.5in]{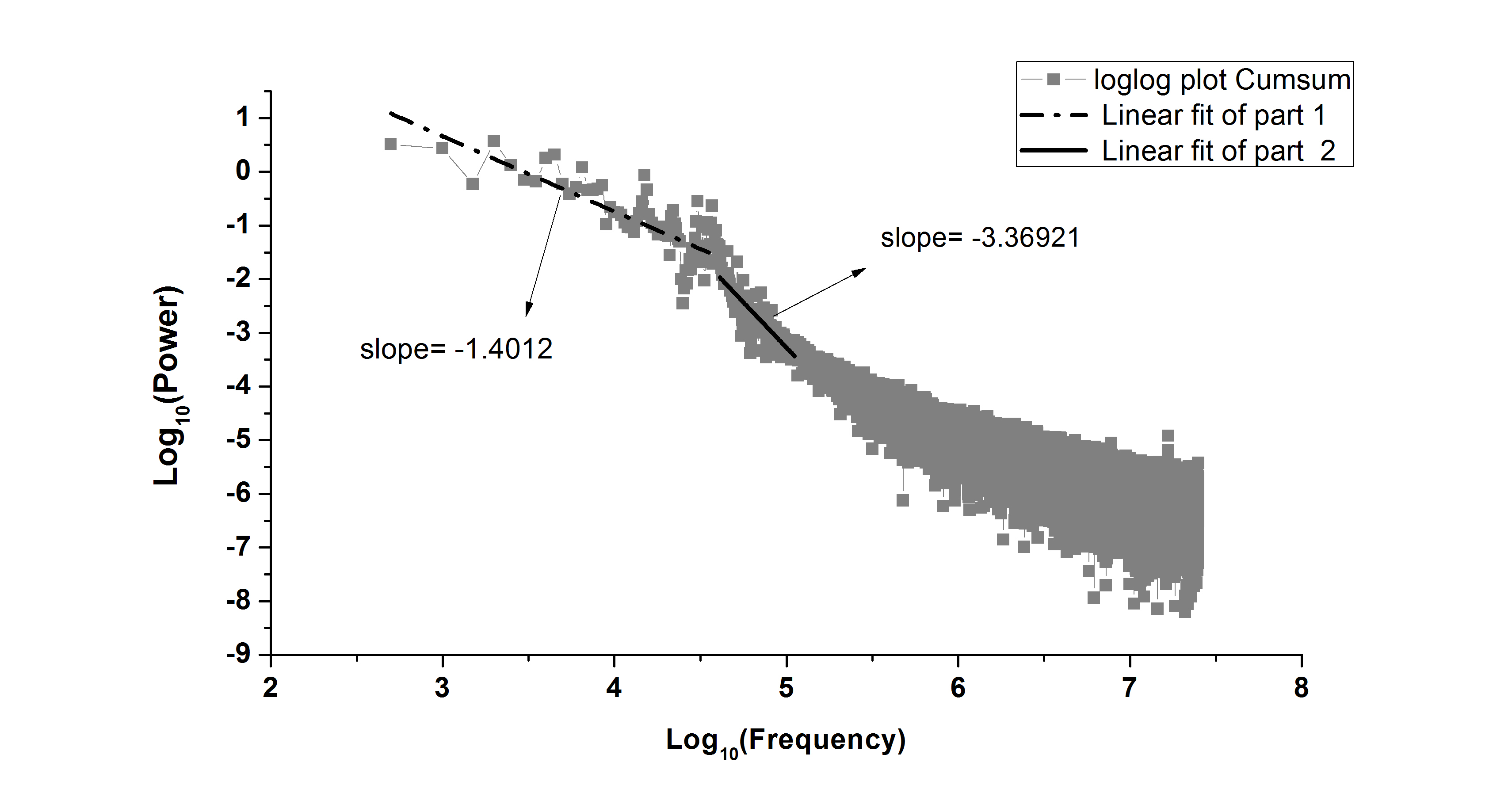}        
    }
    \\
    \subfigure[]
    {
        \includegraphics[height=1.8in,width=3.5in]{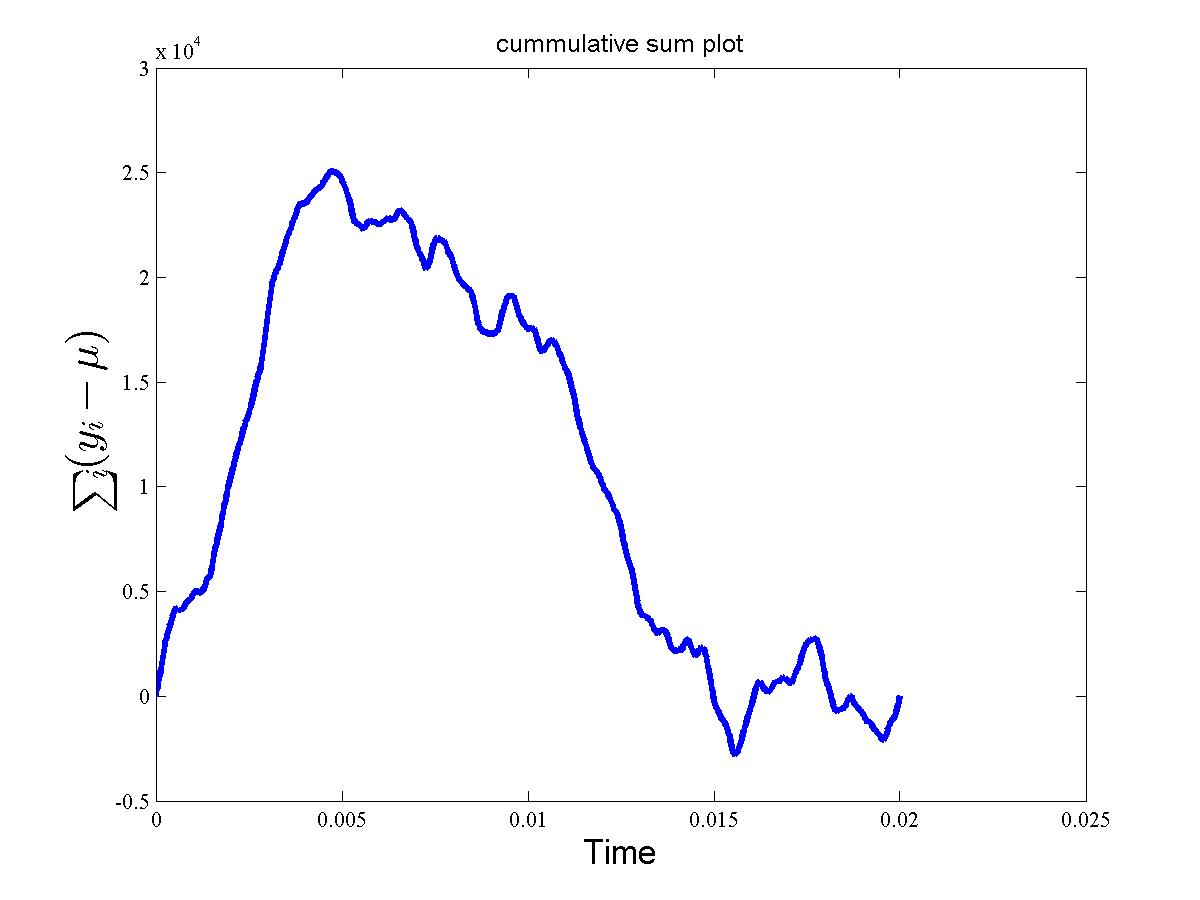}        
    }
    \\
    \subfigure[]
    {
        \includegraphics[height=1.8in,width=3.5in]{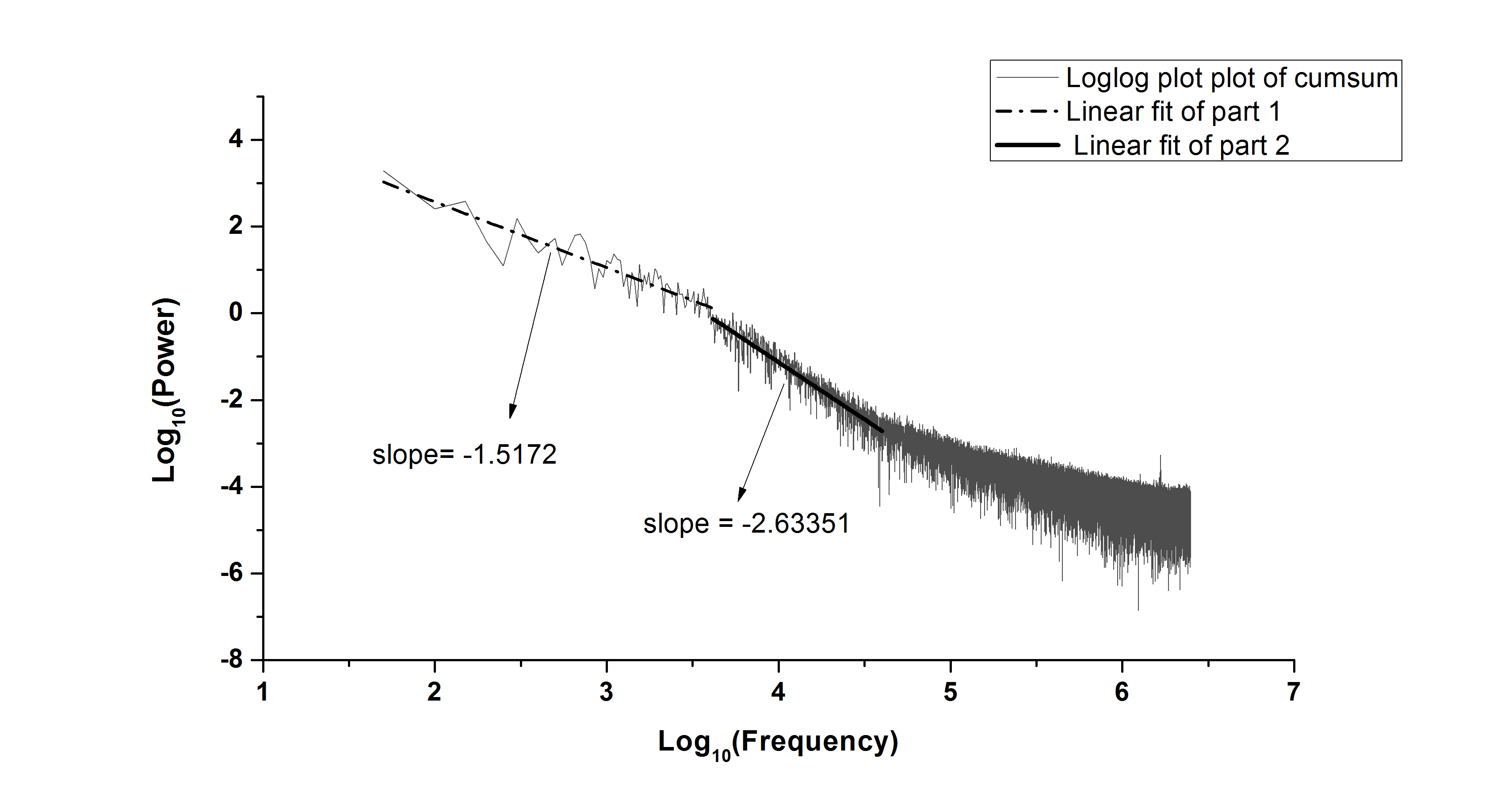}        
    }
    \caption{\label{fig:loglog}(a), (c) Cumulative sum of the fluctuations $y_i = \sum_i(x_i - <x>)$ with $i = 1, ..., N$, for 330 V, 96 gauss and 430 V, 19 gauss respectively and (b), (d) the respective log-log plots of the power spectra, indicating multi-fractal character.}
\end{figure}

Fourier analysis is a widely used tool for testing and quantifying periodic as well as self-similar behavior in a time series. Fractality is quantified through the Hurst exponent ($H$) \cite{Hurst}, which can be related to the power spectrum ($P(k)$) of a fluctuation time series:
\begin{equation*}
P(k)\approx f^{-\alpha}
\end{equation*}
where,
\begin{equation} \label{eq: Fourier_1}
\alpha = 2H+1 ~~ .
\end{equation}
Here, $f$ is the frequency and, $0\leq H\leq 1$, is a signature of fractal behavior \cite{Mandelbrot,WA_Stanley}. We have analysed the recorded plasma discharge potential fluctuations using discrete Fourier transform (DFT), its power spectra is depicted in figures \ref{fig:FFT_1} and \ref{fig:FFT_2}, which illustrate the transition of the system from periodic to chaotic behavior, as the magnetic field is increased. Initially, the system shows a periodic behavior at 330 V and then as the magnetic field is increased, we observe the appearance of higher frequency components. At 430 V, one observes several frequency components, as more energy is pumped into the system on account of the increasing potential. As mentioned before, self-similar character of the potential fluctuations can be quantified from the behavior of the power spectrum. Figure \ref{fig:loglog} depicts the nature of the time series for two characteristic potential values. The power law behavior is evident in the frequency domain.  

The Fourier based methodology of estimating the fractal dimension \cite{Mandelbrot,Feder} is based on the mono-fractal hypothesis, which assumes that the scaling properties are same throughout the time series. This assumption is found to be unrealisable in the present context. As can be observed, the power law coefficient is not uniform in the entire frequency region. This behavior is an indication of the multifractal nature of the signals \cite{FA_1,Panigrahi_express}.

Therefore, it was found to be desirable and efficient to use a  more general analysis which can extract and quantify the multi-fractality. In the process, the inadequacy of the global approach will be pointed out, which will then lead to the wavelet based approach \cite{WA_Panigrahi_2}. Using wavelets the signal is broken down into several subsets, which when analysed separately give the local Hurst exponent describing the local singular behavior and scaling of the signals.

Before proceeding to the wavelet based approach, we make use of rescale range analysis in order to further ascertain the character of the time series.


\subsection*{Rescale Range Analysis}

\begin{figure}[h]
\begin{center}$
\begin{array}{cc}
\includegraphics[scale=0.2]{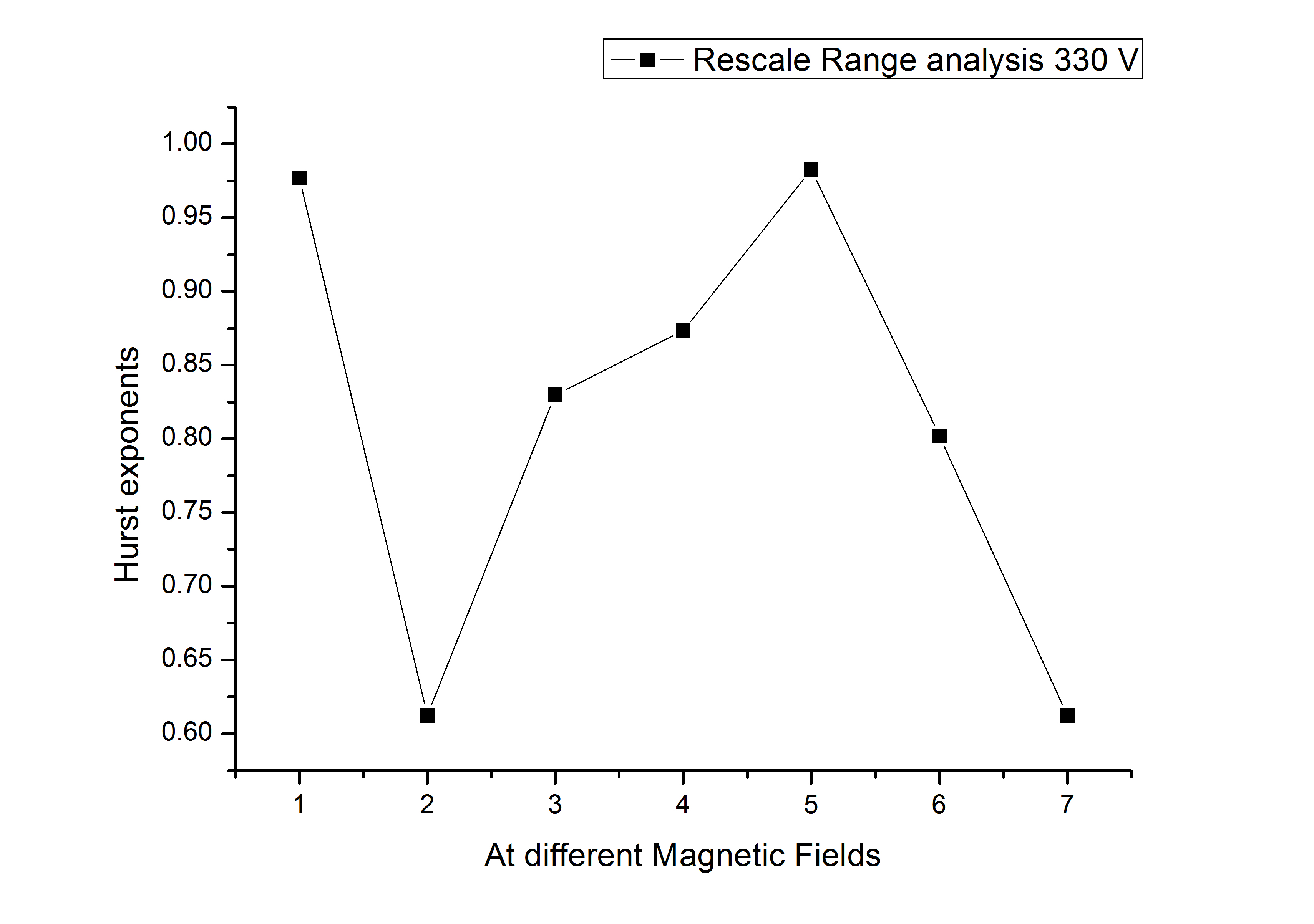}\\
\includegraphics[scale=0.2]{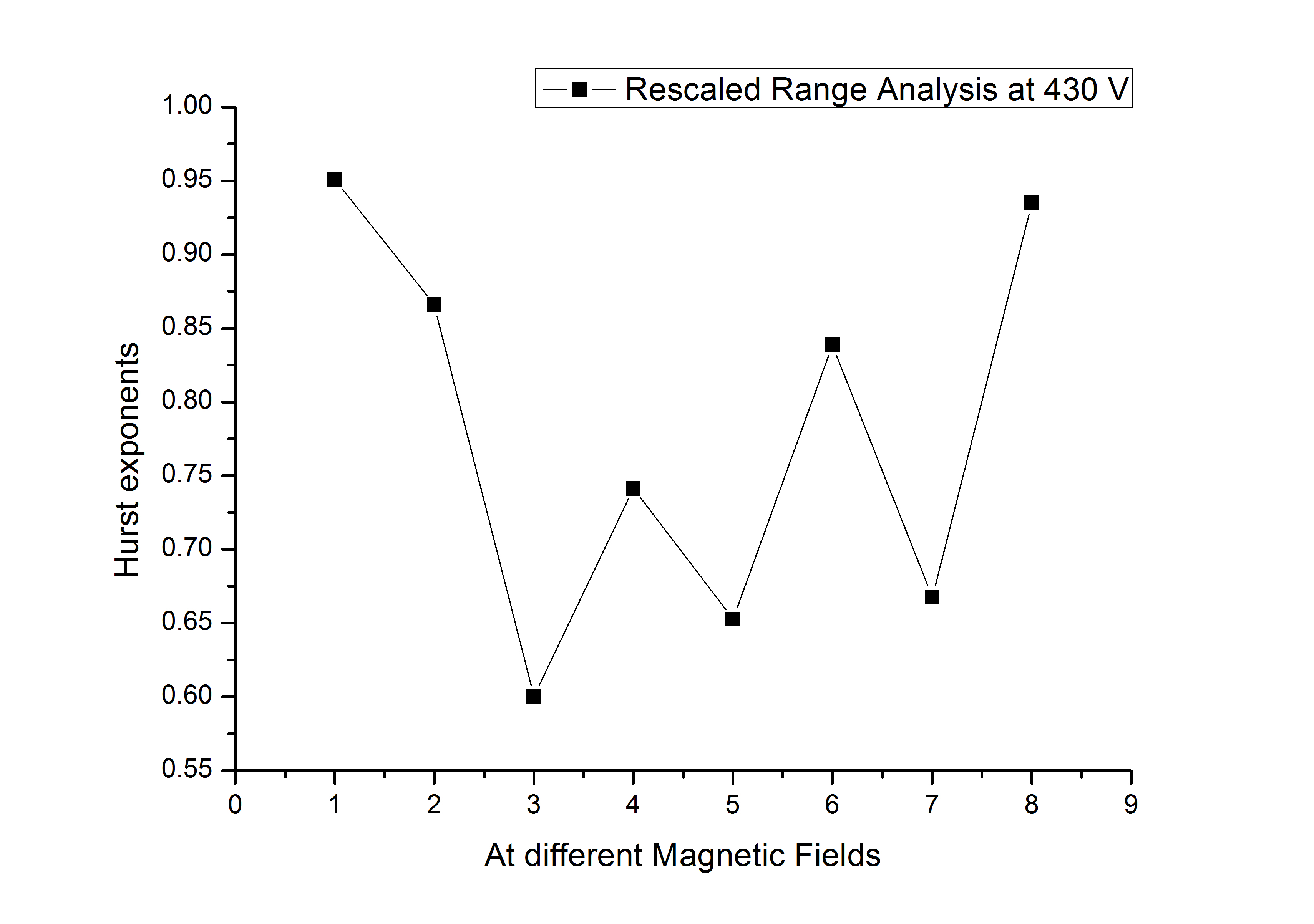} 
\end{array}$
\end{center}
\caption{\label{RRA}Rescale range analysis of the potential time series at (a) 330 V and (b) 430 V show of positive correlation. The ordering of the abscissa is as per Table \ref{Mag}  }
\end{figure}

Rescale range analysis is a technique for providing simultaneously, a measure of variance and the long term correlation in a time series \cite{RRA_1}. Figure \ref{RRA} shows the Hurst exponents calculated using rescale range analysis. Clearly, the Hurst exponents are well above the value of 0.5 showing long range correlations. We also observe that for some threshold value of magnetic field the Hurst exponent changes from $\sim$ 1 to $\sim$ 0.6, showing transition from an extreme fractal towards Brownian motion.

\FloatBarrier

For estimating multi-fractality reliably, one needs to compute other moments, apart from the variance.
One can make use of multi-fractal detrended fluctuation analysis (MFDFA) \cite{WA_Stanley}, for which wavelets can be effectively used for detrending purpose.


\section{Wavelet based analysis} \label{sec:WA}

\subsection*{Wavelet Transform}
A wavelet function is characterized by a translation parameter ($a$) and scale parameter ($s$) \cite{WA_Book}. The wavelet transform of a signal decomposes it into components dependent upon both position and scale. Therefore, the wavelet transform of a time series is localized both in time as well as frequency domains. Unlike the Fourier transform, which gives information solely of the frequency components, the wavelet transform specifies where it occurs in the time series.

\subsection*{Continuous Wavelet Transform (CWT)}
The Continuous Wavelet Transform (CWT) of a function $f(t)$, where $t$ is time, is defined as:
\begin{equation}
W_f(s,a) = \frac{1}{\sqrt{s}}\int_{-\infty}^{+\infty}f(t)\psi^{*}(\frac{t-a}{s})dt
\end{equation}
Here $\psi(t)$ is the mother wavelet and $\psi(t)^*$ its complex conjugate. $\psi_{s,a}(t) = \psi(\frac{t-a}{s})$ yields all the shifted and scaled versions of $\psi(t)$. Changing the wavelet scale via $s$ and translating along the localized time via $a$, the amplitudes of a signal as a function of scale as well as the time can be constructed \cite{WA_Torrence_Compo}. The wavelet should have zero mean and must be localized in time and frequency. The wavelet power spectrum is inferred from $ln |W_f(s,a)|^2$ at various scales $s$ \cite{WA_Panigrahi_1, WA_Panigrahi_2}.

\begin{figure}[!]
\begin{center}$
\begin{array}{cc}
\includegraphics[scale=0.30]{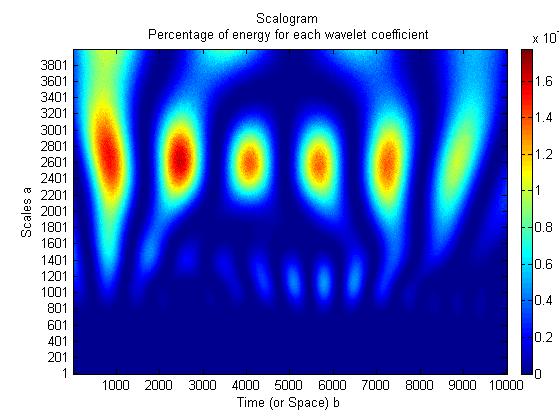}\\
\end{array}$
\end{center}
\caption{\label{Scalograms}Scalograms showing energies at different scales, clearly depicting non-stationary periodic behavior at two different scales.}
\end{figure}

\begin{figure}[!]
    \centering
    \subfigure[]
    {
        \includegraphics[scale=0.15]{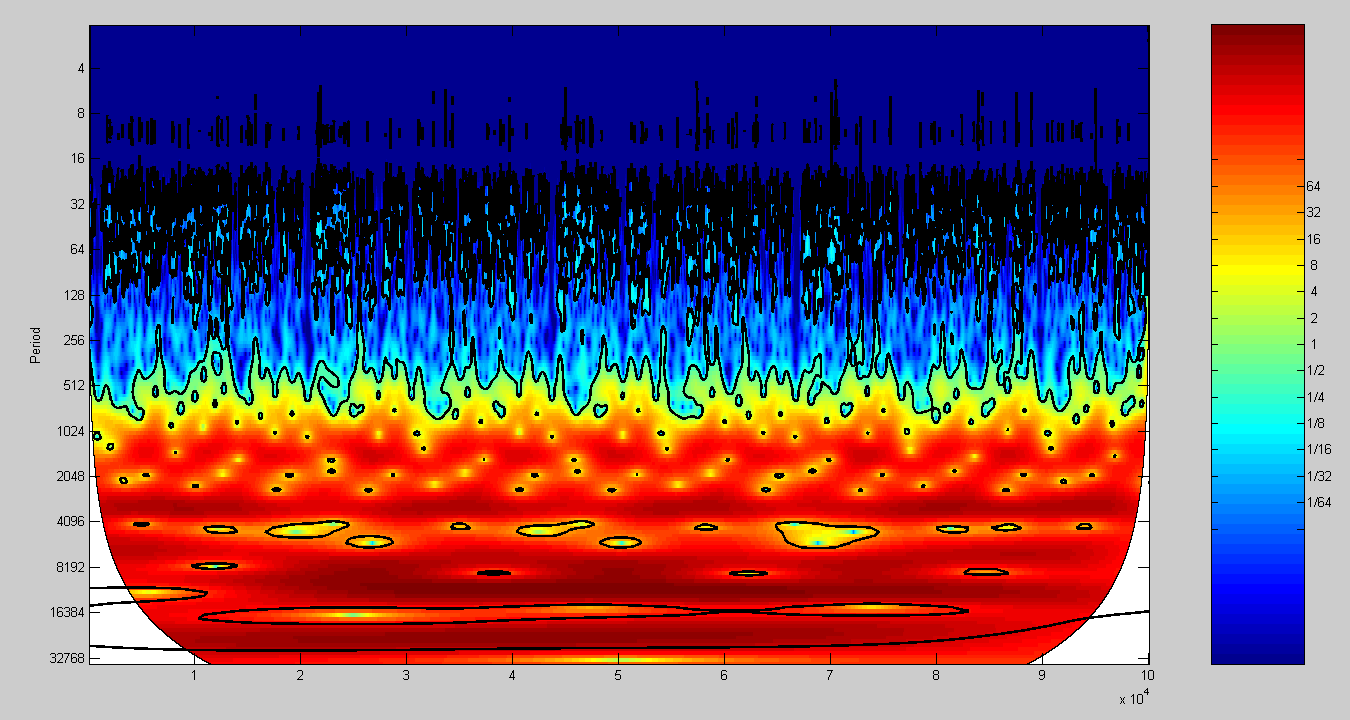}
    }
    \caption{\label{fig:Global}Wavelet power spectrum displaying the cone of influence for the signals at 330 V. The wavelet analysis clearly distinguishes the two dominant periods of the time series with the confidence level 95\%.}
\end{figure}

\begin{figure}[!]
\begin{center}$
\begin{array}{cc}
\includegraphics[scale=0.3]{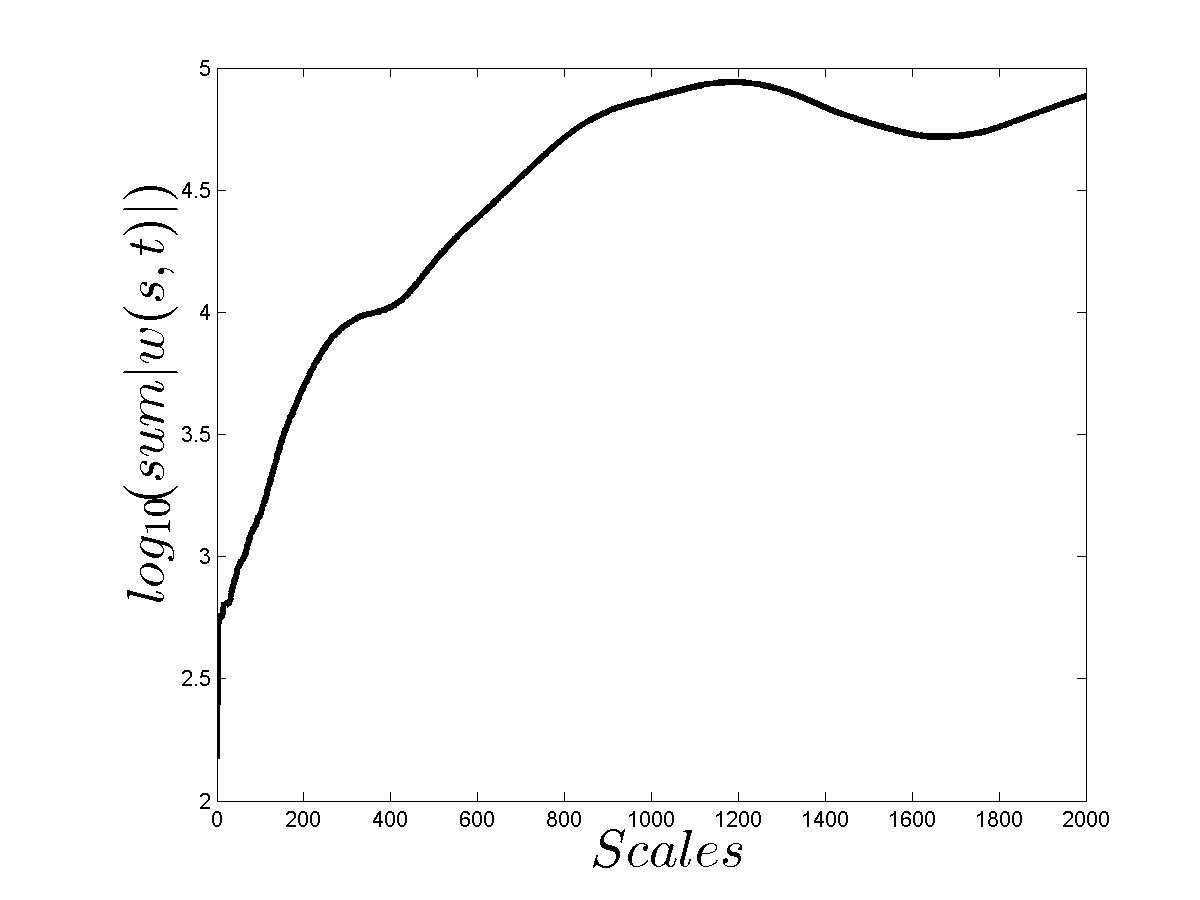}\\
\end{array}$
\end{center}
\caption{\label{Semilog}Semilog plot of wavelet power spectrum summed over all time at different scales, indicating dominant modulations at different scales.}
\end{figure}

The scalogram plot of wavelet coefficients (at different scales) in figure \ref{Scalograms} shows the presence of  periodicity in the time series. Figure \ref{fig:Global} depicts the global wavelet spectrum and average variance with time. It reveals the presence of the periods corresponding to that obtained by DFT. The non-stationary character of the dominant periodic modulations are evident in the scalogram showing the advantage of time-frequency localization of the Morlet wavelets. The cone of influence shows the reliability of the wavelet coefficients in a finite size dataset \cite{Manimaran}. The red regions indicate high confidence level. Semilog plot in figure \ref{Semilog} of wavelet power summed over all time at different scales also shows regular periodic behavior at large scales. 

\begin{figure}[h]
\begin{center}
\includegraphics[scale=0.3]{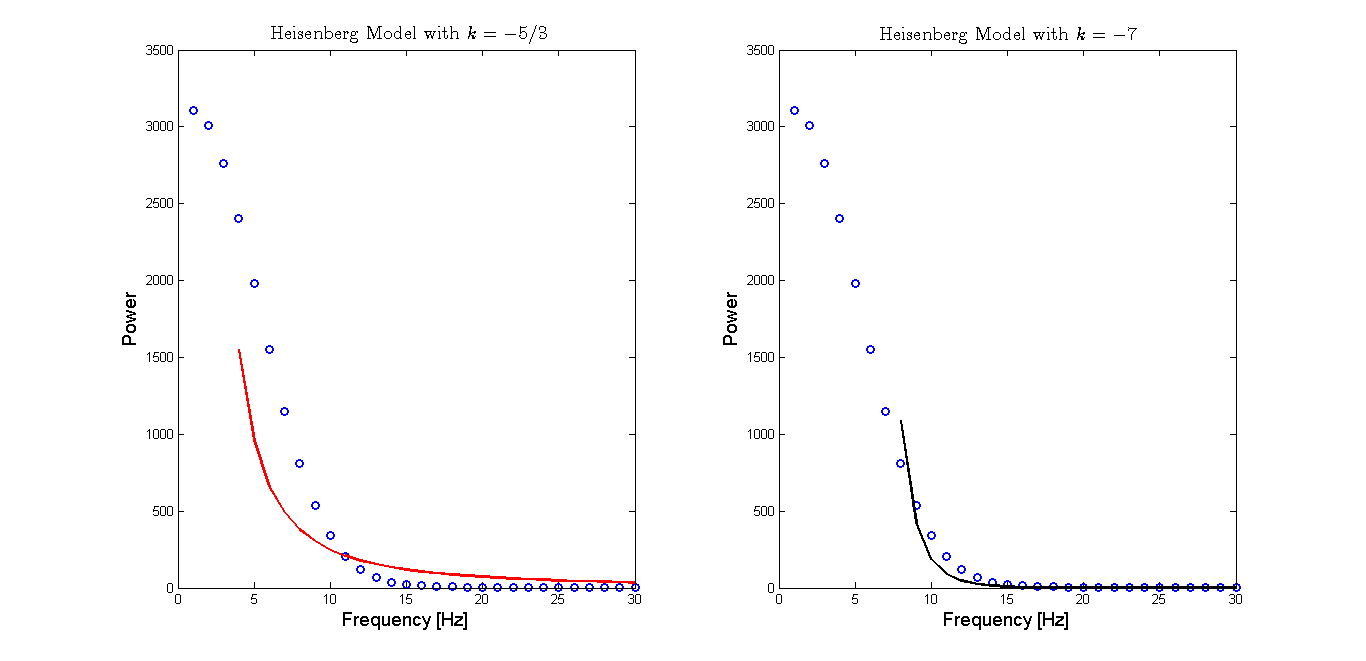}
\caption{\label{fig:Heisenberg}The Hesisenberg fit (solid line) of different exponents. The first fit having slope -5/3,  representing turbulence, while the second fit with slope -7 represents viscous dissipative regime.}
\end{center}
\end{figure}

The presence of turbulence and dissipation can also be inferred through wavelet based approach.A turbulent power spectrum has characteristic slopes as -5/3 in the turbulent \cite{Kol} and -7 in the viscous dissipation regime \cite{Heisen,Farge}. The Heisenberg model exhibits these power laws, where a smooth transition takes place between them. We expect the possibility of neutral turbulence and dissipation in our system, which is revealed with the Heisenberg fit of different exponents shown in figure \ref{fig:Heisenberg} \cite{Heisenberg}. The slopes obtained are -5/3 and -7, which give the turbulent and dissipative nature of the system respectively. Some regions of the system completely match with the fitted curve. Thus, the plasma system has both the properties in suitable parameter domains.

\FloatBarrier


\section{Phase analysis}

The presence of multi-periodic modulations raises the prossibility of their synchronization. The effect of magnetic field on the same is also of deep interest, since magnetic field directly affects the phase of charge carriers. The present section outlines the phase relations obtained from complex CWT, with the complex Morlet function given by,

\begin{equation}
\psi(n) = \frac{1}{\sqrt{(\pi F_b)}}\exp (2iF_cn - \frac{n^2}{F_b})
\end{equation}
Here, $F_b = 1$ and $F_c = 1.5$ are the values of bandwidth parameter and wavelet center frequencies respectively. The phase angle is defined by,
\begin{equation} \label{eq:PA}
\psi(n) = \tan^{-1}(\frac{Im(\psi(n))}{Re(\psi(n))})
\end{equation}

\begin{figure}[h]
    \centering
    \subfigure[]
    {
        \includegraphics[scale=0.1]{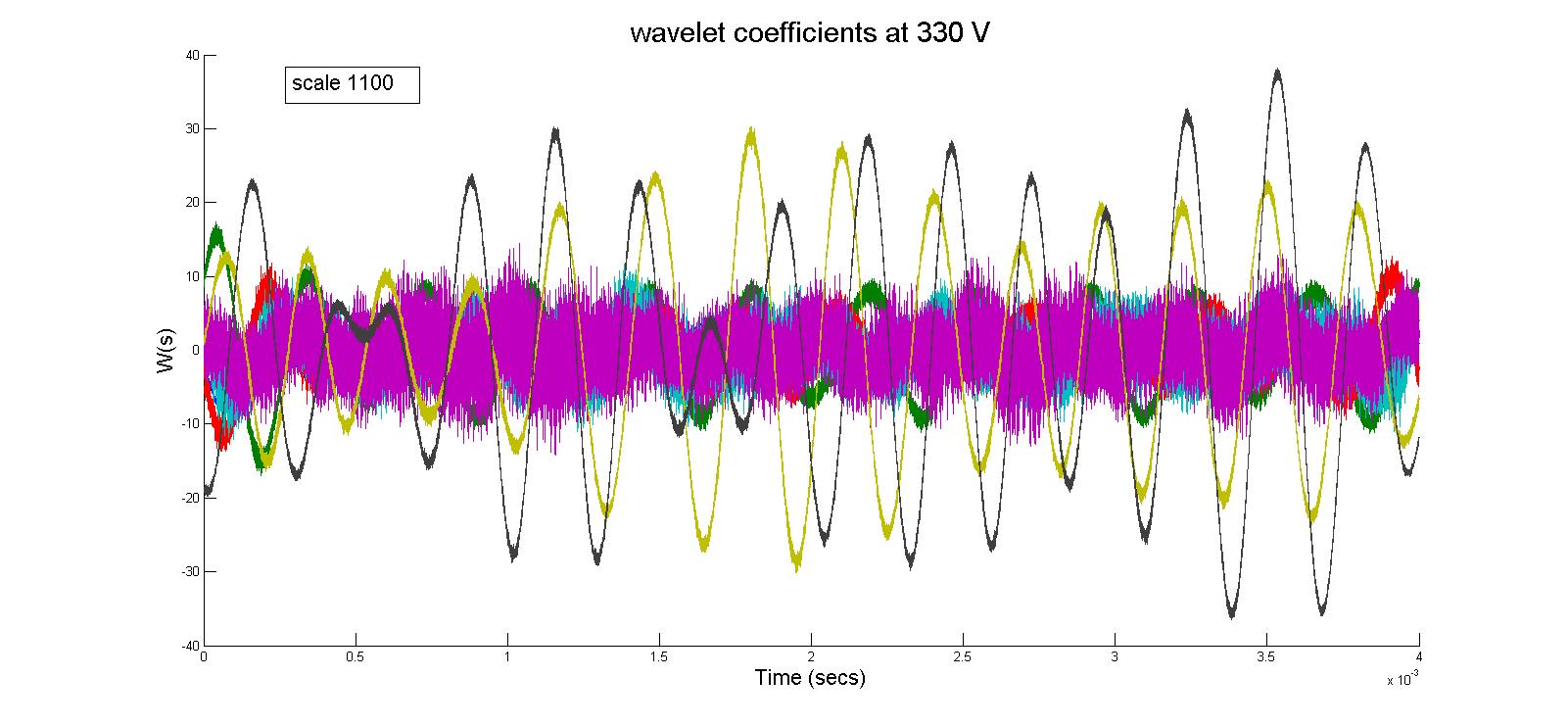} 
    }
    \\
    \subfigure[]
    {
        \includegraphics[scale=0.1]{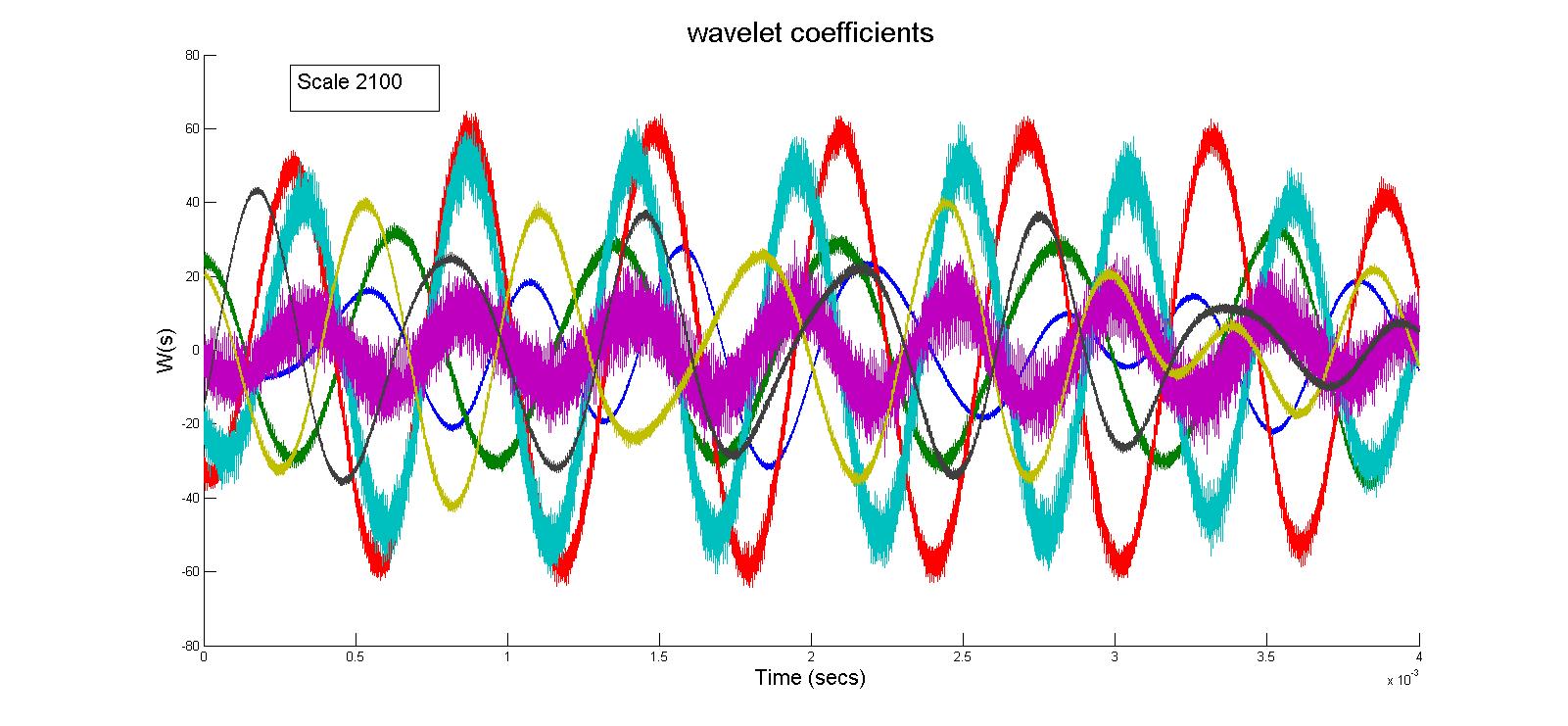}     
    }
    \caption{\label{fig:PA_1}Wavelet coefficients of signal at 330 V, 96 gauss obtained at different scales.}
\end{figure}

\begin{figure}[h]
    \centering
    \subfigure[]
    {
        \includegraphics[scale=0.1]{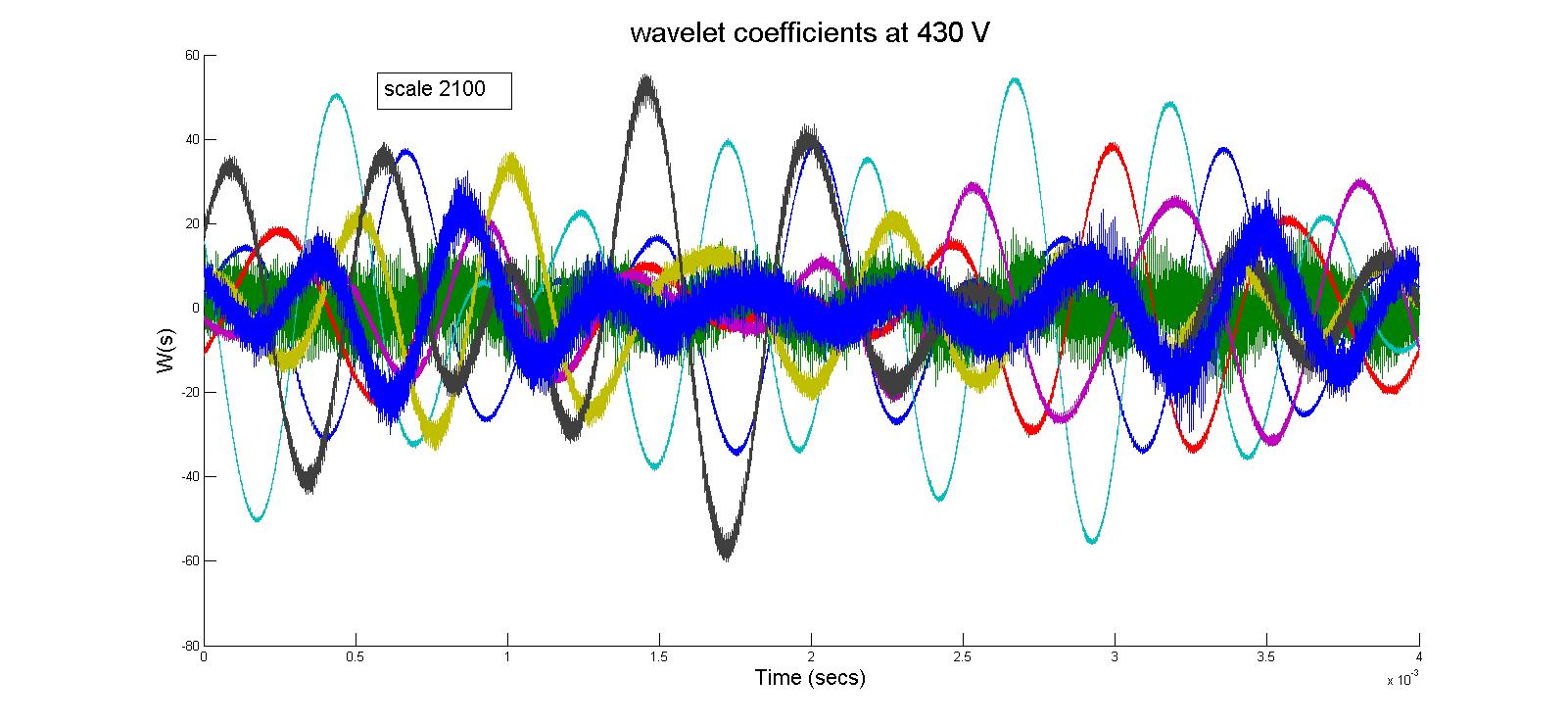} 
    }
    \\
    \subfigure[]
    {
        \includegraphics[scale=0.1]{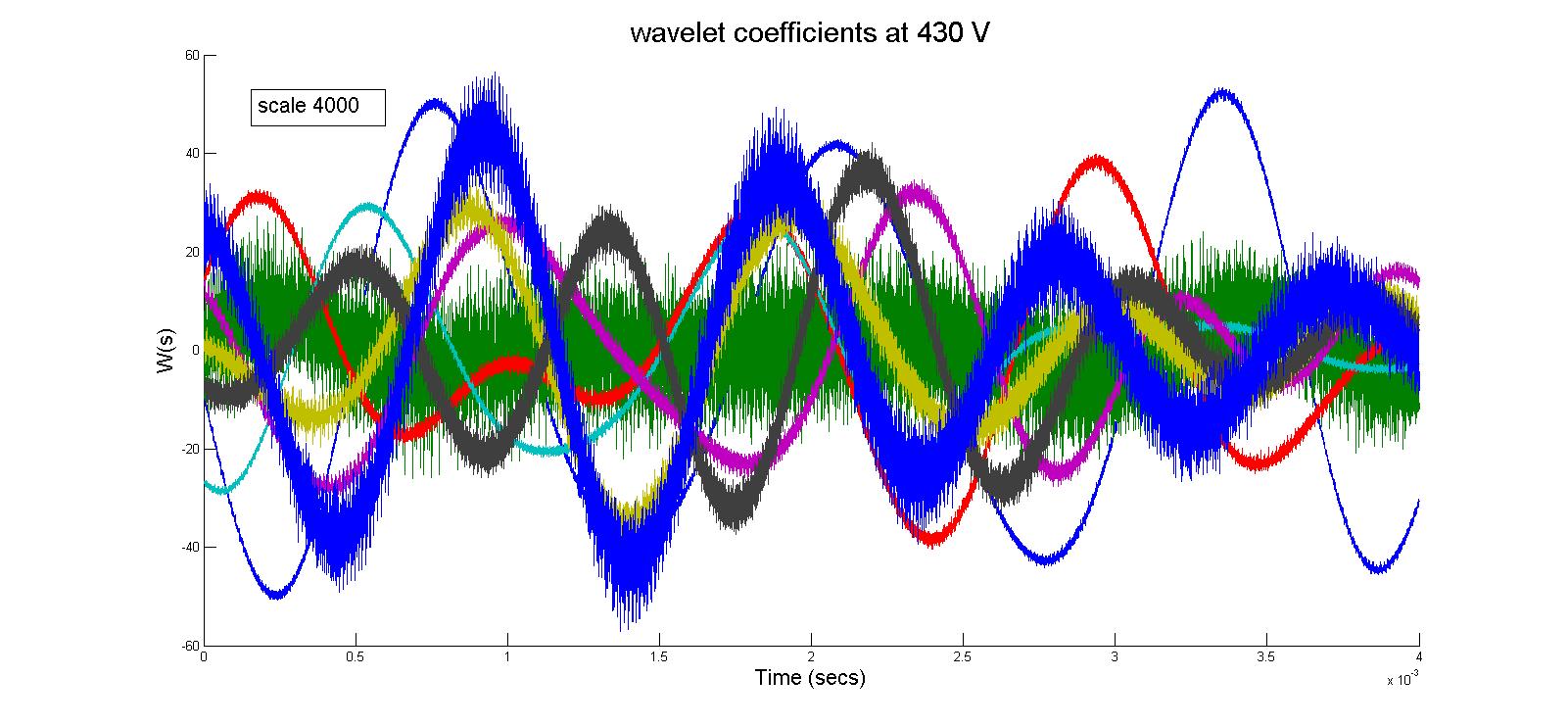}     
    }
    \caption{\label{fig:PA_2}Wavelet coefficients of signal at 430 V, 19 gauss obtained at different scales.}
\end{figure}

The phase relation obtained are plotted in figures \ref{fig:PA_1} and \ref{fig:PA_2} for 330 V and 430 V respectively at different scales. Two dominant periods are clearly visible in figure \ref{fig:PA_1} and four such dominant periods in figure \ref{fig:PA_2}. However, only two of them manifest at a confidence level of 95 $\%$.

\FloatBarrier

\FloatBarrier

\begin{figure}[h]
\centering
\begin{center}
\includegraphics[height=2.7in, width=3.8in]{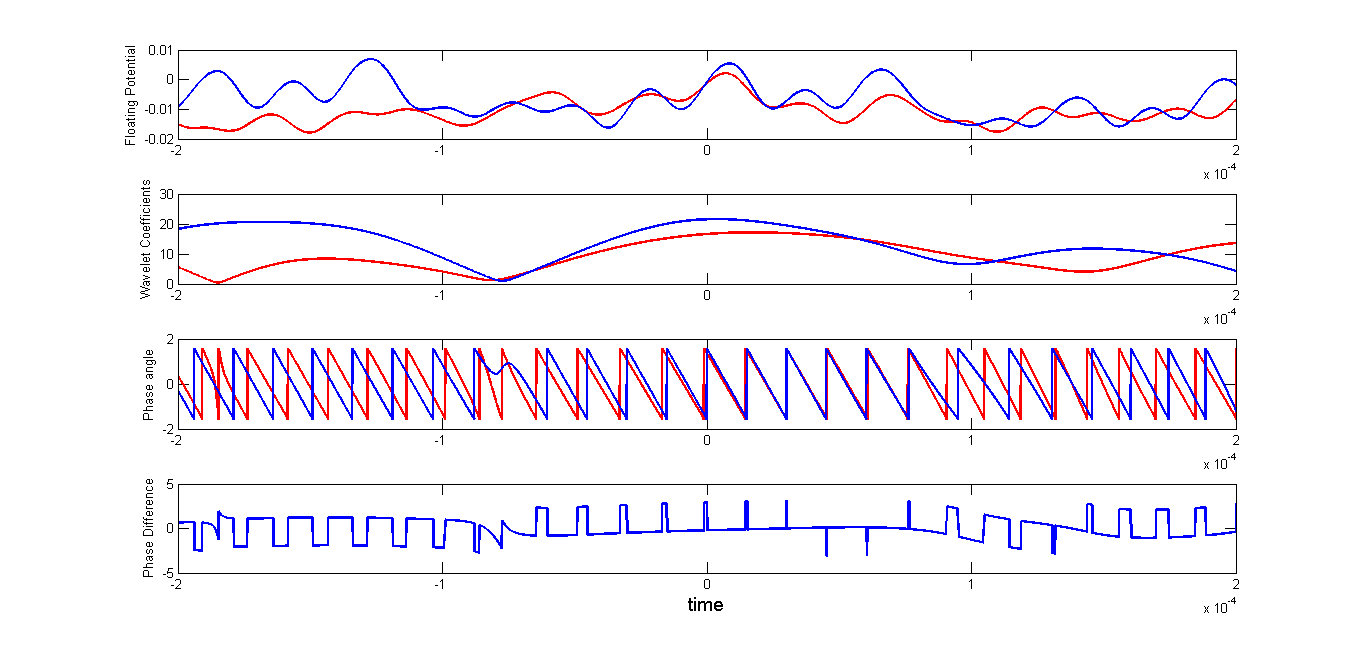}
\end{center}
\caption{\label{fig:PA_3}
Wavelet phase analysis of two floating potential fluctuation signals at 330 V and magnetic fields 87 gauss and 97 gauss respectively. Figure (a) shows the floating potential fluctuations, (b) shows the wavelet coefficients of corresponding signals reconstructed from wavelet spectral analysis, (c) phase angles of the reconstructions, (d) shows the phase difference between the corresponding signals. 
}
\end{figure}

We perform a phase analysis on the periodic modulations to study their synchronization \cite{PA_1}. Figure \ref{fig:PA_3} indicates that the phases change smoothly over time. Initially, the signals are out of phase after which they are locked in more closely. Later, the signals again get out of phase after a small intermittent duration of in-phase dynamics as seen in figure \ref{fig:PA_3}.

\FloatBarrier


\section{Conclusion} \label{sec:Conclusion}

In conclusion, we have studied the floating potential fluctuations in argon DC glow discharge plasma with the application of a magnetic field. We use Fourier and wavelet based techniques to study the non-stationarity character of the time series. The phase space shows periodic nature of the signals. However, as the potential and magnetic field is increased, the system reveals high frequency components.

The non-stationary dynamics is investigated with both discrete and continuous wavelets. Discrete wavelets from the Daubechies family have been effectively used in isolating the local polynomial trends and estimating the multi-fractal nature of the time series. The continuous Morlet wavelet has been handy to identify the periodic and structured variations in the time-frequency domain, as a function of the magnetic field.

We demonstrate the presence of neutral turbulence and dissipation in the plasma system by drawing correspondence with the Heisenberg model, where a slope of -5/3 corresponds to the former, while -7 to the latter.

Further, the effect of magnetic field on the phases of periodic modulations, have been investigated. We illustrate phase synchronization between two time series obtained at two different values of the applied magnetic field, and find that the time series which are initially out-of-phase, go into phase synchronization for an intermittent duration.

Further studies of the dynamics underlying phase synchronization between the time series will be of interest. As time progresses, the system goes through a series of stages where they are either in-phase or out of phase. In this context, the phenomenon of `phase-flip' bifurcation has been recently observed, where dynamical systems coupled with delay, which are initially out-of phase later go into phase as the coupling delay gradually changes with time \cite{PF_Prasad}. 
It would be interesting to build a similar model for the dynamics observed here with variable delay such that it gives rise to such bursts of in-phase and anti-phase synchronization between the time series. This would provide useful insights into the presence and effects of delays on the dynamics of laboratory plasma. The exact nature of dynamics leading to turbulence and dissipation also needs closer scrutiny.


%
%

\end{document}